\documentclass[9pt,twocolumn,twoside]{osajnl}

\journal{pr}

\setboolean{shortarticle}{false}

\usepackage{dcolumn}

\usepackage{graphicx}% Include figure files
\usepackage{dcolumn}% Align table columns on decimal point
\usepackage{bm}% bold math
\usepackage{physics}
\usepackage{color}
\usepackage{ulem}

\usepackage{graphicx}% Include figure files
\usepackage{bm}% bold math
\usepackage{physics}

\title{Deterministic Distribution of Orbital Angular Momentum Multiplexed Continuous-variable Entanglement and Quantum Steering}

\author[1,2,\dag]{Li Zeng}
\author[1,2,3,\dag]{Rong Ma}
\author[1,2]{Hong Wen}
\author[1,2]{Meihong Wang}
\author[1,2]{Jun Liu}
\author[1,2,4]{Zhongzhong Qin}
\author[1,2,*]{Xiaolong Su}

\affil[1]{The State Key Laboratory of Quantum Optics and Quantum Optics Devices,
Institute of Opto-Electronics, Shanxi University, Taiyuan 030006, China}
\affil[2]{Collaborative Innovation Center of Extreme Optics, Shanxi University,
Taiyuan 030006, China}
\affil[3]{College of physics and electronic engineering, Shanxi University,
Taiyuan 030006, China}

\affil[4]{e-mail: zzqin@sxu.edu.cn}
\affil[$^{*}$]{Corresponding author: suxl@sxu.edu.cn}

\date{\today}% It is always \today, today,
\begin{abstract}
Orbital angular momentum (OAM) multiplexing provides an efficient method to improve data-carrying capacity in various quantum communication protocols. It is a precondition to distribute OAM multiplexed quantum resources in quantum channels for implementing quantum communication. However, quantum steering of OAM multiplexed optical fields and the effect of channel noise on OAM multiplexed quantum resources remain unclear. Here, we generate OAM multiplexed continuous-variable (CV) entangled states and distribute them in lossy or noisy channels. We show that the decoherence property of entanglement and quantum steering of the OAM multiplexed states carrying topological charges $l=1$ and $l=2$ are the same as that of the Gaussian mode with $l=0$ in lossy and noisy channels. The sudden death of entanglement and quantum steering of high-order OAM multiplexed states is observed in the presence of excess noise. Our results demonstrate the feasibility to realize high data-carrying capacity quantum information processing by utilizing OAM multiplexed CV entangled states.
\end{abstract}
\setboolean{displaycopyright}{true}

\begin{document}

\maketitle

\maketitle

\section{Introduction} 

Einstein-Podolsky-Rosen (EPR) entanglement plays a crucial role in quantum information processing, such as quantum communication, quantum computation, and quantum precision measurement \cite{EPREntanglement,BraunsteinRMP,KimbleQuanInt,WeedbrookRMP,PhysicsReport}. Besides entanglement, quantum steering, which stands between entanglement \cite{EPREntanglement} and Bell nonlocality \cite{Bell} in the hierarchy of quantum correlations \cite{SteeringRMP}, has been identified as a useful quantum resource. Different from entanglement and Bell nonlocality, quantum steering shows unique asymmetry or even one-way characteristics \cite{WisemanPRL,OneWayNatPhot,ANUexp,OneWayPryde,OneWayGuo,OneWayQin,XiaoY2017,cvdv,WangPRL2020}, and thus allows asymmetric quantum information processing. For example, quantum steering enables one-side device independent quantum key distribution \cite{SteeringQKD,SchnabelOneSidedQKD,PKLamOneSidedQKD}.

Multiplexing provides an efficient method to enhance the data-carrying capability in both classical and quantum communication systems by combining multiple channels into a single channel. By utilizing different degrees of freedom (DOFs) of light, such as wavelength \cite{WaveMultiplexing1,WaveMultiplexing2}, polarization \cite{PolarMultiplexing}, temporal \cite{TimeMultiplexing1,TimeMultiplexing2,TimeMultiplexing3} or spatial \cite{SpatialMultiplexing1,SpatialMultiplexing2} modes, different types of multiplexing can be realized. Orbital angular momentum (OAM) of light \cite{AllenOAM} has also been found to be an attractive DOF to realize multiplexing due to its infinite range of possibly achievable topological charges \cite{OAMMultiplexing1,OAMMultiplexing2}. OAM has found applications in discrete-variable quantum information processing, such as high dimensional OAM entanglement generation \cite{ZeilingerPNAS}, and 18-qubit entanglement with six photons’ three DOFs including OAM \cite{Lu18modes}. 

\begin{figure*}[t]
\begin{center}
\includegraphics[width=13cm]{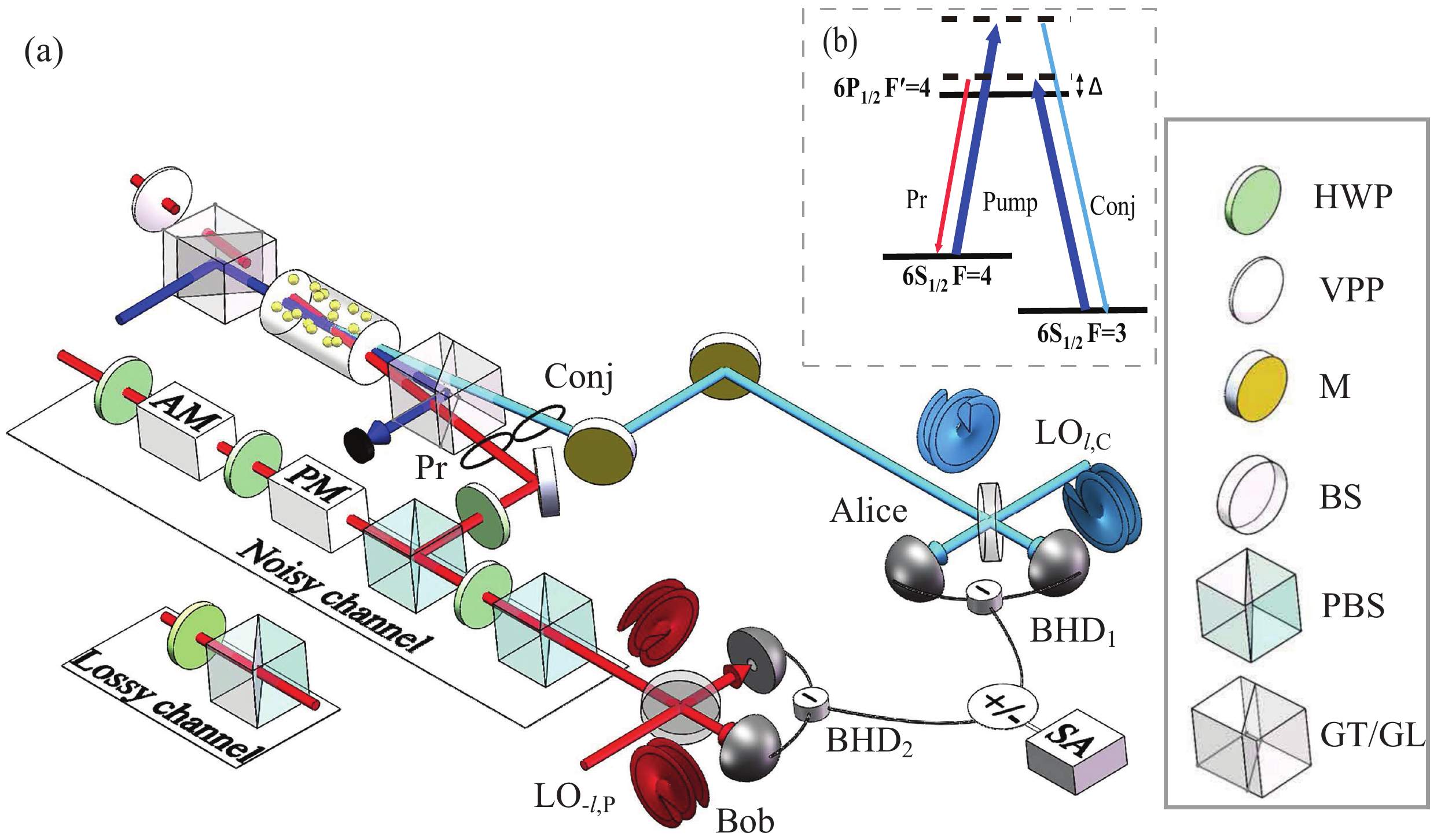}% Here is how to import EPS art
\end{center}
\caption{(a) Experimental setup for the generation and distribution of OAM multiplexed CV quantum entanglement and steering in a lossy or noisy channel. Pr: probe beam; Conj: conjugate beam; LO\(_{-l, P}\) and LO\(_{l, C}\): local oscillators of Pr and Conj fields; AM: amplitude modulator; PM: phase modulator; GL: Glan-laser polarizer; GT: Glan-Thompson polarizer; PBS: polarization beam splitter; HWP: half-wave plate; VPP: vortex phase plate; M: mirror; BS: 50:50 beam splitter; BHD$_{1}$, BHD$_{2}$: balanced homodyne detectors; SA: spectrum analyzer. (b) Double-$\Lambda$ energy level structure for the FWM process in cesium vapor cell. $\Delta$: one-photon detuning.}
\end{figure*}

Four-wave mixing (FWM) process in warm alkali vapor cell has found a wide range of applications \cite{QinLight,QinPRL,EntangledImages,JingHexaEnt,OAMFWM2008,SU11,PooserOptica}. Especially, spatial-multi-mode advantage of the FWM process, attributed to its cavity-free configuration, makes it an ideal optical parametric amplifier to generate entangled images \cite{EntangledImages} and reconfigurable multipartite entanglement \cite{JingHexaEnt}. Quantum correlated twin beams carrying OAM were generated based on the FWM process in rubidium vapor \cite{OAMFWM2008}. OAM multiplexed bipartite and multipartite CV entangled states have been generated based on the FWM process \cite{JingBiOAM,JingTriOAM,JingHexaOAM}. Furthermore, OAM multiplexed deterministic all-optical quantum teleportation has also been demonstrated by utilizing OAM multiplexed bipartite CV entangled state generated from the FWM process \cite{JingQuanTele}. To enhance the data-carrying capacity in quantum communication based on OAM multiplexed CV entangled states, it is essential to distribute them in lossy and noisy quantum channels towards practical applications. The distribution of weak coherent field and single photons carrying OAM in fiber, free space, and underwater have been experimentally investigated \cite{OAMDistribution1,OAMDistribution2,OAMDistribution3}. However, it remains unclear whether the quantum entanglement and steering of OAM multiplexed CV entangled states are more sensitive to loss and noise than commonly used Gaussian mode with $l=0$.

Here, we present the deterministic distribution of OAM multiplexed CV quantum entanglement and steering in lossy and noisy channels. In the experiment, the OAM multiplexed entangled fields are generated deterministically based on the FWM process in warm cesium vapor and distributed deterministically in quantum channels. We show that the CV entangled states carrying topological charges $l=1$ and $l=2$ are as robust against loss as Gaussian mode with $l=0$. Sudden death of entanglement and quantum steering of high-order OAM multiplexed CV entangled state is observed in the presence of noise. Our results pave the way for applying OAM multiplexed CV entanglement and quantum steering in high data-carrying capacity quantum communication.

\medskip
\section{The principle and experimental setup} 

Figure 1(a) shows the schematic of experimental setup, and Fig. 1(b) shows the double-$\Lambda$ energy level structure used for the FWM process, which is formed from the $^{133}$Cs $D_{1}$ line with an excited level ($6P_{1/2}, F'=4$) and two ground levels ($6S_{1/2}, F=3$ and $F=4$). The pump beam is about 1.6 GHz  blue detuned from $6S_{1/2}, F=3\rightarrow6P_{1/2}, F'=4$ transition, and the probe beam is 9.2 GHz red shifted relative to the pump beam. The pump and probe beams are combined by a Glan-laser (GL) polarizer and then cross each other in the center of the cesium vapor cell at an angle of 6 mrad \cite{MaOL}. The gain of the FWM process is around 3 with a pump power of 240 mW and a probe power of 3 $\mu$W. By injecting the probe beam carrying topological charge $l$ of OAM mode, conjugate beam carrying topological charge $-l$ of OAM mode is generated on the other side of the pump, which satisfies OAM conservation in the FWM process. The topological charge of OAM mode $l=1$ or $l=2$ is added to the probe beam by passing it through a vortex phase plate (VPP). The pump beam is filtered out by using a Glan-Thompson (GT) polarizer with an  extinction ratio of $10^{5}$:1 after the vapor cell. 

The Conj field is kept by Alice, while the Pr field is distributed to a remote quantum node owned by Bob through a lossy or noisy channel. The lossy channel is simulated by a half-wave plate (HWP) and a polarization beam splitter (PBS). The noisy channel is modeled by combining the Pr field with an auxiliary beam at a PBS followed by a HWP and a PBS. The auxiliary beam carries the same frequency and topological charge with the Pr field, and is modulated by an amplitude modulator and a phase modulator with white noise \cite{WangPRL2020}. To characterize the OAM multiplexed CV entangled state, its covariance matrix (CM) is experimentally measured by utilizing two sets of balanced homodyne detectors (BHDs). In order to extract the CV quadrature information carried by the OAM mode with a topological charge $l$, local oscillator (LO) with opposite topological charge $-l$ is required. In our experiment, the spatially mode-matched LO beams used in the BHDs are obtained from a second set of FWM process which is around 5 mm above the first set of FWM process in the same vapor cell \cite{EntangledImages}. More details of the experimental parameters can be found in Appendix B. 
\begin{figure*}[htbp]
\label{Loss}
\begin{center}
\includegraphics[width=16 cm]{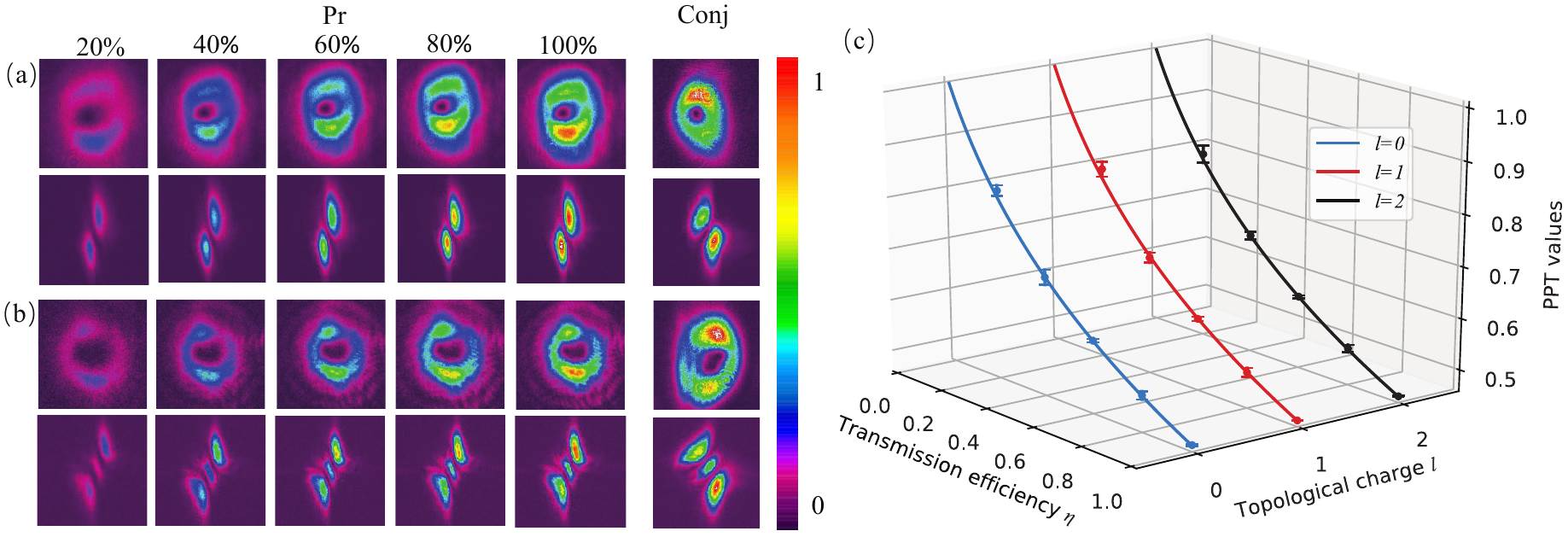}% Here is how to import EPS 
\end{center}
\caption{\label{Loss} (a), (b) Beam patterns of the OAM multiplexed CV entanglement for  \(l=1\) and  \(l=2\) in a lossy channel and corresponding transmitted patterns through a tilted lens. (c) Dependence of PPT values of the OAM multiplexed CV entanglement on transmission efficiency $\eta$ for \(l=0\), \(l=1\) and \(l=2\) in lossy channels. The PPT value is 0.46$\pm$0.01 at $\eta=1$. Curves and data points show theoretical predictions and experimental results, respectively. Error bars of experimental data represent one standard deviation and are obtained based on the statistics of the measured data.}
\end{figure*}

The Hamiltonian of the OAM multiplexed FWM process can be expressed as \cite{JingBiOAM}:
\begin{equation}
\hat{H}=\sum_{l}i\hbar\gamma_{l}\hat{a}^{\dagger}_{l,P}\hat{a}^{\dagger}_{-l,C}+h.c.
\end{equation}
where \(\gamma_{l}\), which is defined as the interaction strength of each OAM pair, is proportional to the pump power. \(\hat{a}^{\dagger}_{l,P}\) and \(\hat{a}^{\dagger}_{-l,C}\) are the creation operators related to OAM modes of the Pr and Conj fields, respectively. Since the pump beam does not carry OAM (\(l=0\)), the topological charges of the Pr and Conj fields are opposite due to OAM conservation in the FWM process. The output state of the OAM multiplexed FWM process is:

\begin{equation}
\ket{\Psi}_{out}=\ket{\Psi}_{-l}\otimes\cdots\otimes\ket{\Psi}_{0}\otimes\cdots\otimes\ket{\Psi}_{l}
\end{equation}
where \(\ket{\Psi}_{l}=\ket{\psi_{l,P},\psi_{-l,C}}\) presents a series of independent OAM multiplexed CV EPR entangled state generated in the FWM process. $\ket{\psi_{l,P}}$ and $\ket{\psi_{-l,C}}$ represent Pr field carrying topological charge $l$ and Conj field carrying topological charge $-l$ , respectively.

All Gaussian properties of the CV entangled state \(\ket{\Psi}_{l}\) can be determined by the covariance matrix $\sigma_{AB}$ with the matrix element $\sigma _{ij}=\langle \hat{\xi}_{i}\hat{\xi}_{j}+\hat{\xi}_{j}\hat{\xi}_{i}\rangle /2-\langle \hat{\xi}_{i}\rangle \langle \hat{\xi}_{j}\rangle $, where $\hat{\xi}\equiv (\hat{X}_{-l,C}, \hat{Y}_{-l,C}, \hat{X}_{l,P}, \hat{Y}_{l,P})^{T}$, $\hat{X}=\hat{a}+\hat{a}^{\dag}$ and $\hat{Y}=(\hat{a}-\hat{a}^{\dag})/i$ represent amplitude and phase quadratures of the Conj and Pr fields, respectively. The covariance matrix of the OAM multiplexed entangled state after distribution in a lossy or noisy channel is as following

\begin{equation}
\sigma_{AB,\delta,\eta}=\left(\begin{array}{cccc}
V_{a} & 0 &  V_{c}  & 0 \\
0 & V_{a} & 0 & -V_{c} \\
V_{c}  & 0 & V_{b} & 0 \\
0 & -V_{c} & 0 & V_{b}
\end{array}
\right)
\end{equation}
with $V_{a}=\frac{V+V^{\prime}}{2}$, $V_{b}=\eta\frac{V+V^{\prime}}{2}+(1-\eta)(1+\delta)$ and $V_{c}=\sqrt{\eta} \frac{V^{\prime}-V}{2}$. $V$ and $V^{\prime}$ represent the variances of squeezed and anti-squeezed quadratures of the optical mode, respectively. The derivation of the covariance matrix can be found in Appendix A. For pure squeezed states, $VV^{\prime}=1$. While $VV^{\prime}>1$ after pure squeezed states suffer from loss or noise. $\eta$ and $\delta$ represent transmission efficiency and excess noise of the quantum channel, respectively. We have $\delta=0$ in a lossy channel, while we have $\delta>0$ in a noisy channel. The submatrices $\sigma _{A}=V_{a} I$ and $\sigma _{B}=V_{b} I$, with $I$ as the identity matrix, are the covariance matrices corresponding to the states of Alice's and Bob's subsystems, respectively. The CV entangled state is a symmetric state and an asymmetric state when $\sigma _{A}=\sigma _{B}$  and $\sigma _{A}\neq \sigma _{B}$, respectively.

\begin{figure}[htbp]
\begin{center}
\label{Noise}
\includegraphics[width=8 cm]{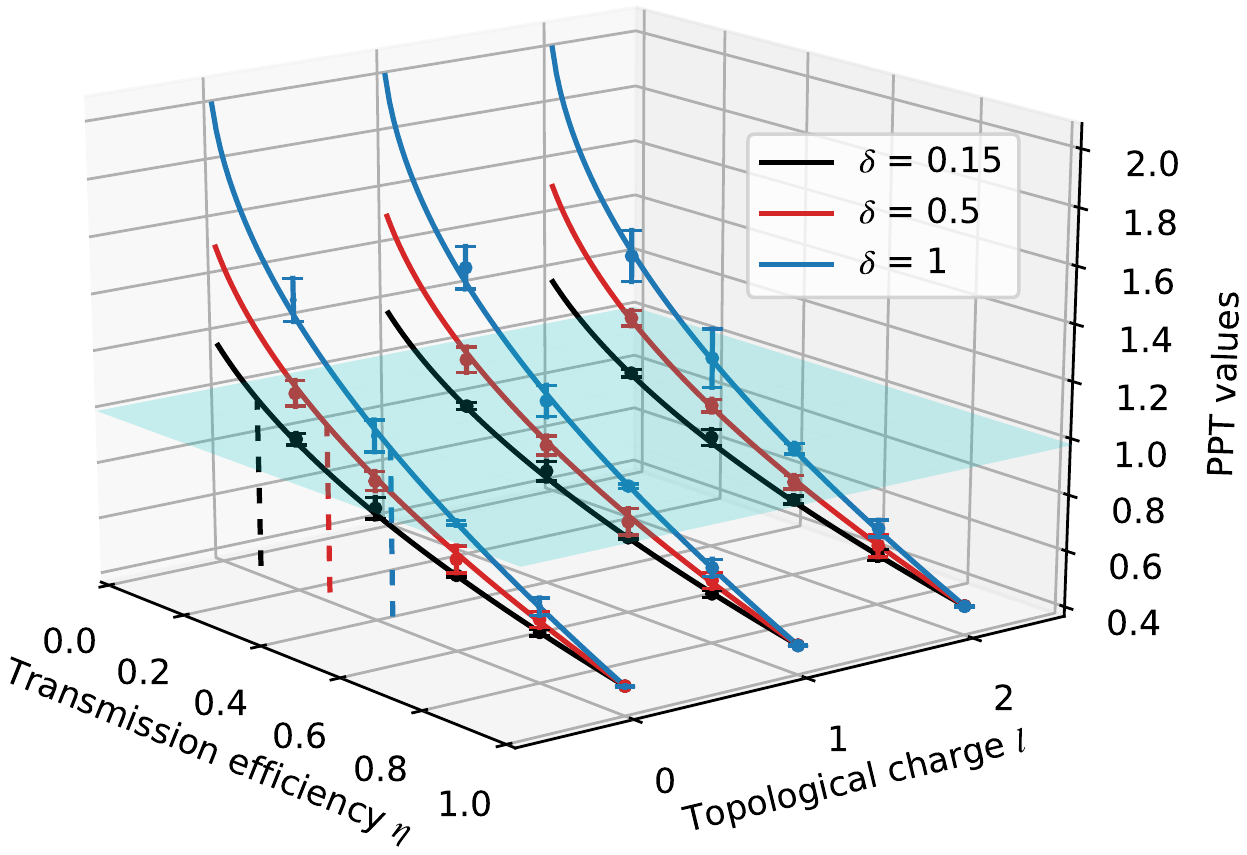}% Here is how to import EPS art
\end{center}
\caption{\label{Noise} Dependence of PPT values of the OAM multiplexed CV entanglement on transmission efficiency $\eta$ for \(l=0\), \(l=1\) and \(l=2\) in noisy channels. Three different amounts of excess noise \(\delta=0.15\) (black), \(\delta=0.5\) (red), and \(\delta=1\) (blue) are compared. The light blue plane shows the boundary for sudden death of entanglement where PPT value equals to 1. The three vertical dashed lines indicate corresponding transmission efficiencies where entanglement starts to disappear. Curves and data points show theoretical predictions and experimental results, respectively. Error bars of experimental data represent one standard deviation and are obtained based on the statistics of the measured data.}
\end{figure}

\begin{figure*}[htbp]
\begin{center}
\includegraphics[width=16 cm]{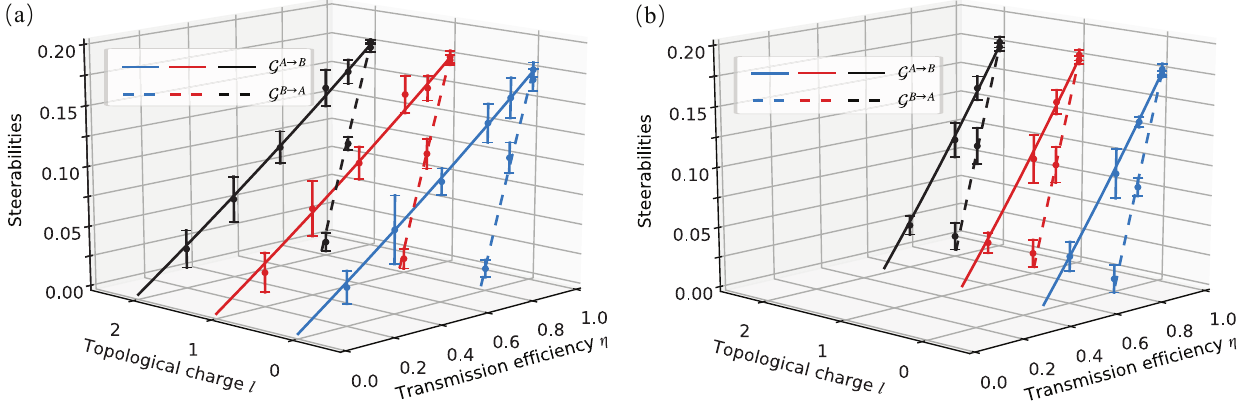}% Here is how to import EPS art
\end{center}
\caption{\label{Steering} Quantum steerabilities of OAM multiplexed CV entangled state distributed in a lossy (a) or noisy (b) channel. The excess noise  shown in (b) is $\delta=0.15$. Solid and dashed curves show theoretical predictions of $\mathcal{G}^{A\rightarrow B}$ and $\mathcal{G}^{B\rightarrow A}$, respectively. Data points show experimental results. Error bars of experimental data represent one standard deviation and are obtained based on the statistics of the measured data.}
\end{figure*}

The Peres-Horodecki criterion of positivity under partial transpose (PPT) criterion is a sufficient and necessary criterion to characterize the entanglement of CV bipartite entanglement \cite{PPTCriterion}. If the smallest symplectic eigenvalue \(\nu\) of the partially transposed covariance matrix is smaller than 1, bipartite entanglement exists. Otherwise, it's a separable state. Furthermore, smaller \(\nu\) represents stronger entanglement.

Quantum steering for bipartite Gaussian states of CV systems
can be quantified by \cite{QuantificationPRL}
\begin{equation}
\mathcal{G}^{A\rightarrow B}(\sigma _{AB})=%
\mbox{$\max\big\{0,\,
\frac12 \ln {\frac{\det \sigma_{A}}{\det \sigma_{AB}}}\big\}$},
\end{equation}
where $\mathcal{G}%
^{A\rightarrow B}(\sigma _{AB})>0$ represents that Alice has the ability to
steer Bob's state. Similarly, we have
\begin{equation}
\mathcal{G}^{B\rightarrow A}(\sigma _{AB})=%
\mbox{$\max\big\{0,\,
\frac12 \ln {\frac{\det \sigma_{B}}{\det \sigma_{AB}}}\big\}$},
\end{equation}
which represents Bob's ability to steer Alice's state. From the expressions of $\mathcal{G}%
^{A\rightarrow B}(\sigma_{AB})$ and $\mathcal{G}^{B\rightarrow A}(\sigma_{AB})$, it can be seen that Alice and Bob have the same steerability if $\det \sigma _{A}=\det \sigma _{B}$ is satisfied; i.e., the bipartite Gaussian state is a symmetric state. If the state is an asymmetric state, the steerabilities of Alice and Bob will be different.

\section{Results} 

To verify the OAM property of the optical fields, we measure the spatial beam patterns of quantum states  \(\ket{\Psi}_{1}\) and \(\ket{\Psi}_{2}\) transmitted through a lossy channel, which are shown in the top rows of Fig. \ref{Loss}(a) and \ref{Loss}(b), respectively. It is obvious that the Pr and Conj fields are both Laguerre-Gaussian beams. To infer their topological charges, they are passed through a tilted lens and imaged on a camera. As shown in the bottom rows of Fig. \ref{Loss}(a) and \ref{Loss}(b), the number of dark stripes gives the number of the topological charge and the direction gives its sign \cite{TopologicalCharge}. As the transmission efficiency of the Pr field decreases, its optical intensity also decreases, while its topological charge remains unchanged. Additional beam patterns of the Pr and Conj fields can be found in Appendix D.

The covariance matrices of the OAM multiplexed entangled states are reconstructed by measuring the noise variances of the amplitude and phase quadratures of the Conj and Pr fields $\Delta ^{2}\hat{X}_{-l,C}$, $\Delta ^{2}\hat{Y}_{-l,C}$, $\Delta ^{2}\hat{X}_{l,P}$, and $\Delta ^{2}\hat{Y}_{l,P}$, as well as their correlation variances of amplitude and phase quadratures $\Delta ^{2}(\hat{X}_{l,P}-\hat{X}_{-l,C}) $, and $\Delta ^{2}(\hat{P}_{l,P}+\hat{P}_{-l,C}) $, respectively. Details about the measurement of the covariance matrices can be found in Appendix C. Based on the covariance matrix of each OAM multiplexed entangled state at different loss and noise levels, its quantum entanglement and quantum steering characteristics are evaluated experimentally. 

Fig. \ref{Loss}(c) shows the dependence of PPT values of the CV bipartite entangled state carrying different topological charges on the transmission efficiency of the Pr field. The correlation and anti-correlation levels of the initial CV entangled states carrying topological charges \(l=0\), \(l=1\), and \(l=2\) are all around $-3.3$ dB and 6.1 dB, which correspond to $V=0.47$ and $V^{\prime}=4.11$, respectively. The entanglement between the Pr and Conj fields degrades as the transmission efficiency decreases. However, the entanglement is robust against loss, i.e., it always exists until the transmission efficiency reaches 0. It is obvious that the CV bipartite entangled state carrying topological charges \(l=1\), \(l=2\) are as robust to loss as their Gaussian counterpart \(l=0\).

Figure \ref{Noise} shows the dependence of PPT values of the CV bipartite entangled state carrying different topological charges in noisy channels. Compared with the results in Fig. \ref{Loss}(c), the entanglement disappears at a certain transmission efficiency of the Pr field in the presence of excess noise, which demonstrates the sudden death of CV quantum entanglement. Furthermore, the higher the excess noise is, the sooner entanglement disappears. The transmission efficiencies where entanglement starts to disappear are $\eta=0.10, 0.28$ and 0.44, respectively, for the excess noise $\delta=0.15, 0.5$ and 1 in the units of shot-noise level (SNL). We show that OAM multiplexed CV entangled states carrying high order topological charges \(l=1\), \(l=2\) exhibit the same decoherence tendency as their Gaussian counterpart \(l=0\) in noisy channels. 

The dependence of steerabilities $\mathcal{G}^{A\rightarrow B}$ and $\mathcal{G}^{B\rightarrow A}$ on the transmission efficiency $\eta$ and topological charge $l$ in lossy and noisy channels are shown in Fig. \ref{Steering} (a) and  Fig. \ref{Steering} (b), respectively. In a lossy channel, the steerabilities for both directions always decrease when the transmission efficiency decreases. One-way steering is observed in the region of $0<\eta <0.72$ for OAM multiplexed CV entangled state carrying different topological charges \(l=0\), \(l=1\), and \(l=2\). In a noisy channel, where the excess noise \(\delta=0.15\) (in the units of SNL) exists, the steerabilities $\mathcal{G}^{A\rightarrow B}$ and $\mathcal{G}^{B\rightarrow A}$ are lower than those in the lossy channel \cite{OneWayQin,npjDeng}. Futhermore, Alice loses its steerability in the region of $0<\eta <0.45$, while Bob loses its steerability in the region of $0<\eta <0.75$, which confirms sudden death of quantum steering in a noisy channel  \cite{npjDeng}. It is worth noting that the CV entangled state carrying topological charges \(l=1\), \(l=2\) has the same steerabilities as their counterpart \(l=0\).

\section{Conclusion} 

The distribution of OAM multiplexed CV entanglement and quantum steering in quantum channels with homogeneous loss and noise, such as fiber channels, are experimentally simulated in our work. There are also other quantum channels with inhomogeneous loss and noise, such as atmospheric turbulence and diffraction. Recently, it has been shown that other optical fields carrying OAM, such as vector beams, are turbulence-resilient in atmospheric turbulence \cite{VectorBeam}. Thus it is worthwhile to investigate the turbulence-resilient characteristics of OAM multiplexed CV quantum entanglement and steering, which have the potential to substantially improve the quantum communication distance and fidelity. 

In summary, we experimentally demonstrate quantum steering of OAM multiplexed optical fields and investigate the distribution of OAM multiplexed CV entanglement and quantum steering in quantum channels. We show that the decoherence property of CV entanglement and quantum steering of the OAM multiplexed optical fields carrying topological charges $l=1$ and $l=2$ are the same as that of the counterpart Gaussian mode with $l=0$ in lossy and noisy channels. The sudden death of entanglement and quantum steering of high-order OAM multiplexed optical fields is observed in the presence of excess noise. Our results demonstrate the feasibility to improve the quantum communication capacity in practical quantum channels by utilizing OAM multiplexed CV entanglement and quantum steering. 

\begin{figure*}[htbp]
\begin{center}
\includegraphics[width=15cm]{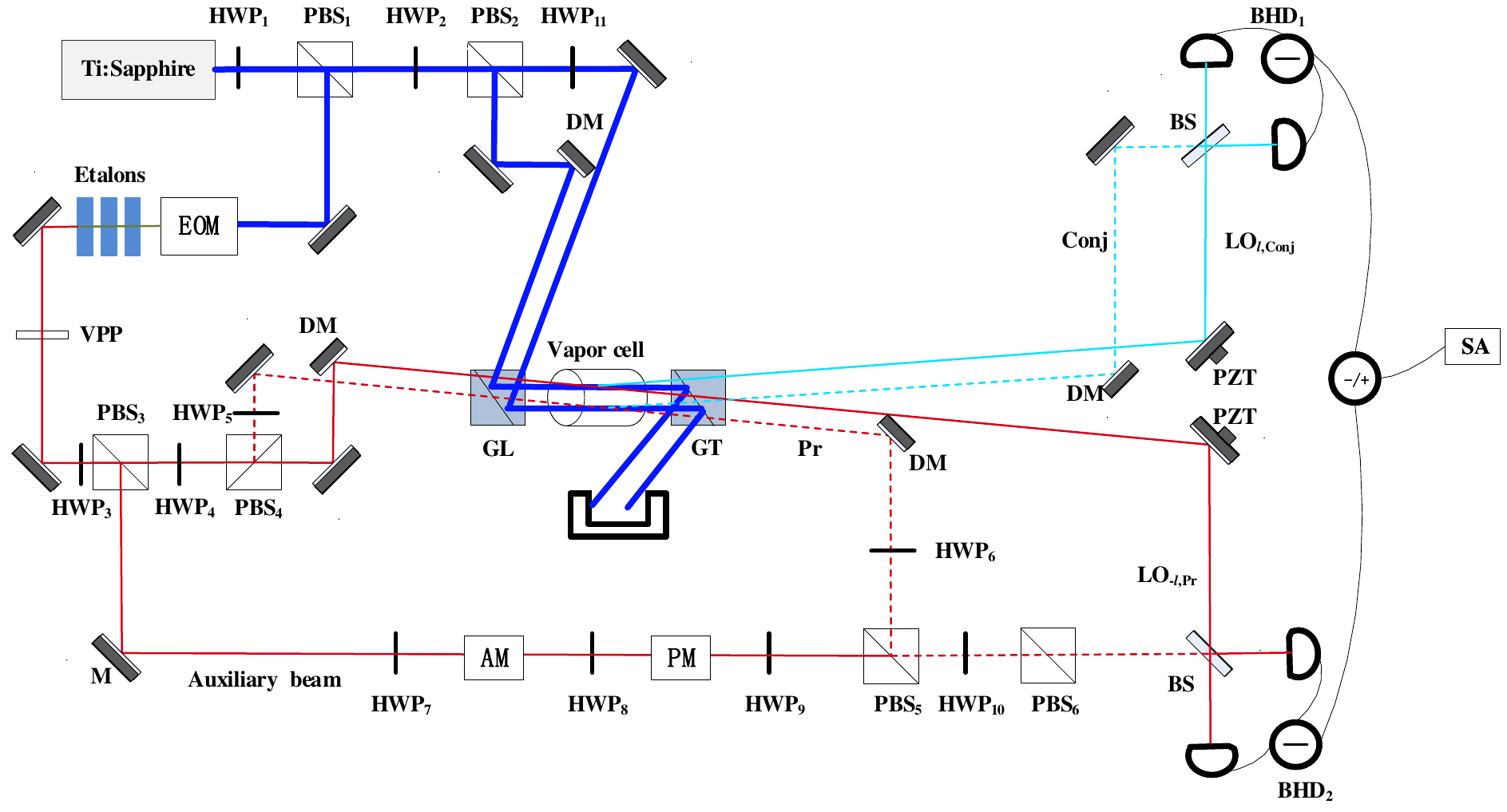}% Here is how to import EPS art
\end{center}
\caption{Detailed experimental schematic for distributing OAM multiplexed CV entanglement in a noisy channel. The lossy channel is realized by blocking the  auxiliary beam. D-shaped mirrors are utilized to combine or separate light beams with small distances. HWP: half-wave plate; PBS: polarization beam splitter; EOM: electro-optic modulator; VPP: vortex phase plate; GL: Glan-laser polarizer; GT: Glan-Thompson polarizer; Pr: probe beam; Conj: conjugate beam; AM: amplitude modulator; PM: phase modulator; M: mirror; DM: D-shaped mirror; PZT: piezoelectric ceramics; BS: 50:50 beam splitter; BHD: balanced homodyne detector; SA: spectrum analyzer.}
\end{figure*}

\begin{figure*}[htbp]
\begin{center}
\includegraphics[width=18cm]{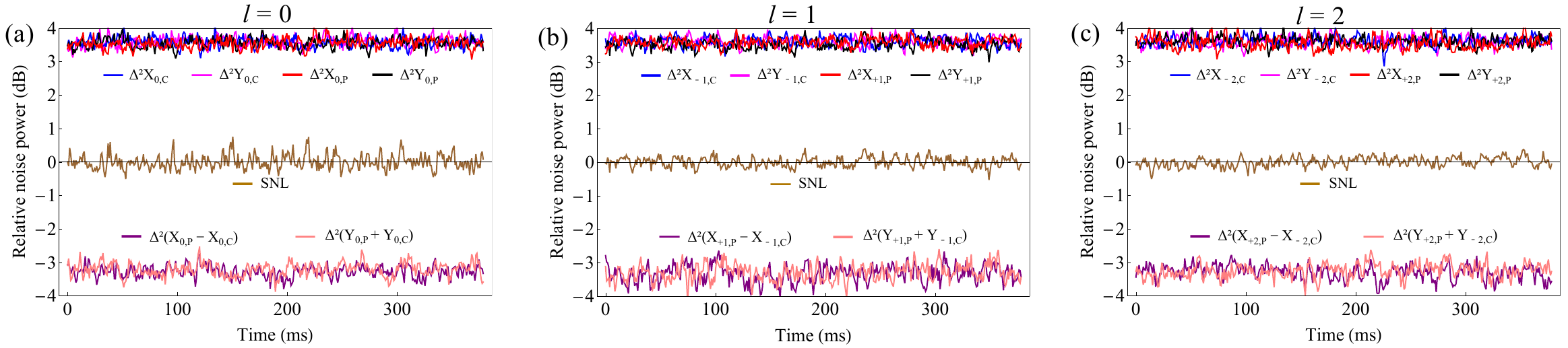}% Here is how to import EPS art
\end{center}
\caption{The measured quantum correlation noises for initially generated OAM multiplexed CV entangled states carrying topological charge \(l=0\) (a), \(l=1\) (b) and \(l=2\) (c), respectively. Brown curve at 0 dB shows the SNL. Other six curves show the noise variances of $\Delta ^{2}\hat{X}_{-l,C}$, $\Delta ^{2}\hat{Y}_{-l,C}$, $\Delta ^{2}\hat{X}_{l,P}$, $\Delta ^{2}\hat{Y}_{l,P}$, as well as noise variances of their joint amplitude or phase quadrature $\Delta ^{2}(\hat{X}_{l,P}-\hat{X}_{-l,C}) $, and $\Delta ^{2}(\hat{P}_{l,P}+\hat{P}_{-l,C}) $. These six curves are all normalized to the same SNL. All the measurements are performed at 1.2 MHz. The electronic noise of the BHDs and background noise from leaked pump fields are subtracted from the SNL and signals, respectively.}
\end{figure*}

\begin{figure*}[htbp]
\begin{center}
\includegraphics[width=18cm]{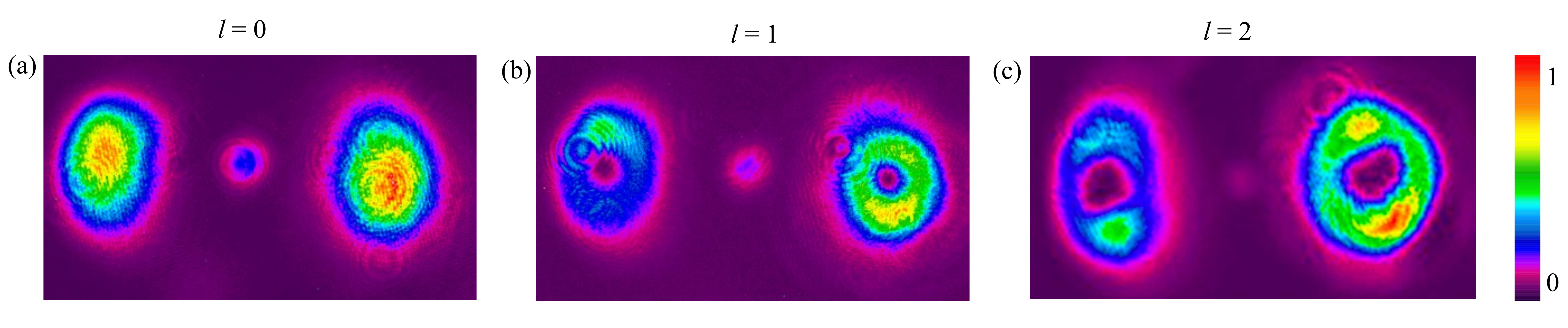}% Here is how to import EPS art
\end{center}
\caption{The images of OAM modes of the Pr beam and Conj beam generated from FWM process. (a) \(l=0\); (b) \(l=1\); (c) \(l=2\). From left to right: Conj beam, leaked pump beam, and Pr beam.}
\end{figure*}

\section*{APPENDIX A: Theoretical model}
The Hamiltonian of the orbital-angular momentum (OAM) multiplexed four-wave mixing (FWM) process can be expressed as:
\begin{equation}
\hat{H}=\sum_{l}i\hbar k_{l}\hat{a}_{0,Pump}\hat{a}_{0,Pump}\hat{a}^{\dagger}_{l,P}\hat{a}^{\dagger}_{-l,C}e^{i\theta}+h.c.
\end{equation}
where $\hat{a}_{0,Pump}$, $\hat{a}^{\dagger}_{l,P}$, and $\hat{a}^{\dagger}_{-l,C}$ are the annihilation operator of the pump field, the creation operators related to OAM modes of the Pr and Conj fields, respectively. $\theta$ is the phase of the pump field. The first term represents the process in which two pump photons are converted into a Pr photon and a Conj photon. $k_{l}$ is interaction strength of each OAM pair. Since the pump beam does not carry OAM (\(l=0\)), the topological charges of the Pr and Conj fields are opposite due to OAM conservation in the FWM process. The second term h.c. is the Hermitian conjugate of the first term, and represents the reversed process in which a Pr photon and a Conj photon are converted to two pump photons. 

The pump field is much stronger than the Pr and Conj fields in the FWM process, so it can be regarded as classical field. By combing the intensity of the pump field $|\alpha_{Pump}|^{2}$ with $k_{l}$, i.e. $|\alpha_{Pump}|^{2}k_{l}=\gamma_{l}$, and taking $\theta=0$, the Hamiltonian can be simplified as \cite{JingBiOAM}:
\begin{equation}
\hat{H}=\sum_{l}i\hbar\gamma_{l}\hat{a}^{\dagger}_{l,P}\hat{a}^{\dagger}_{-l,C}+h.c.
\end{equation}

The output state of the OAM multiplexed FWM process is as following:
\begin{equation}
\ket{\Psi}_{out}=\ket{\Psi}_{-l}\otimes\cdots\otimes\ket{\Psi}_{0}\otimes\cdots\otimes\ket{\Psi}_{l}
\end{equation}
where $\ket{\Psi}_{l}=\hat{S}(\gamma_{l})\ket{vac}=\ket{\psi_{l,P},\psi_{-l,C}}$ means the EPR entangled state with Pr and Conj fields carrying topological charge $l$ and $-l$, respectively. $\ket{vac}$ presents vacuum state, and $\hat{S}(\gamma_{l})=e^{\gamma_{l}(\hat{a}^{\dagger}_{l,P}\hat{a}^{\dagger}_{-l,C}-\hat{a}_{l,P}\hat{a}_{-l,C})}$ presents two-mode squeezing operator for the vacuum state carrying OAM topological charge $l$. It is obvious that a series of CV EPR entangled states carrying independent topological charges are generated in the FWM process, i.e., OAM multiplexing is realized.

All Gaussian properties of the CV Gaussian entangled state \(\ket{\Psi}_{l}\) can be determined by its covariance matrix $\sigma_{AB}$, with the matrix element $\sigma _{ij}=\langle \hat{\xi}_{i}\hat{\xi}_{j}+\hat{\xi%
}_{j}\hat{\xi}_{i}\rangle /2-\langle \hat{\xi}_{i}\rangle \langle \hat{\xi}%
_{j}\rangle $, where $\hat{\xi}\equiv (\hat{X}_{-l,C}, \hat{Y}_{-l,C}, \hat{X}_{l,P}, \hat{Y}_{l,P})^{T}$, $\hat{X}=\hat{a}+\hat{a}^{\dag}$ and $\hat{Y}=(\hat{a}-\hat{a}^{\dag})/i$ represent amplitude and phase quadratures of the Conj and Pr fields. 

The covariance matrix of CV bipartite entangled state can be written as:
\begin{equation}
\sigma_{AB}=\left(\begin{array}{cccc}
\frac{V+V^{\prime}}{2} & 0 & \frac{V^{\prime}-V}{2}  & 0 \\
0 & \frac{V+V^{\prime}}{2} & 0 & \frac{V-V^{\prime}}{2}  \\
\frac{V^{\prime}-V}{2}  & 0 & \frac{V+V^{\prime}}{2} & 0 \\
0 & \frac{V-V^{\prime}}{2}  & 0 & \frac{V+V^{\prime}}{2}
\end{array}
\right) =\left(
\begin{array}{cc}
\frac{V+V^{\prime}}{2} I & \frac{V^{\prime}-V}{2}  Z \\
\frac{V^{\prime}-V}{2}  Z & \frac{V+V^{\prime}}{2} I%
\end{array}%
\right) 
\end{equation}
where $V$ and $V^{\prime}$ represent the variances of squeezed and anti-squeezed quadratures of the optical mode, respectively. $I$ and $Z$ are
the Pauli matrices:
\begin{equation}
I=\left(
\begin{array}{cc}
1 & 0 \\
0 & 1%
\end{array}%
\right) ,\quad Z=\left(
\begin{array}{cc}
1 & 0 \\
0 & -1%
\end{array}%
\right) 
\end{equation}

Then we consider the distribution of CV entangled state \(\ket{\Psi}_{l}\) in a lossy and noisy channel. Let $\hat{a}_{l, P}$ and $\hat{a}_{-l, C}$ represent the annihilation operators of the Pr and Conj fields, respectively. After the Pr field $\hat{a}_{l, P}$ is distributed in a lossy channel, it becomes $\hat{a}^{\prime}_{l, P}=\sqrt{\eta }\hat{a}_{l, P}+\sqrt{1-\eta }\hat{\mu}$, where $\hat{\mu}$ represents vacuum state with variance of 1. Similarly, the Pr field becomes $\hat{a}^{\prime}_{l, P}=\sqrt{\eta }\hat{a}_{l, P}+\sqrt{1-\eta }(\hat{\epsilon}+\hat{\mu})$ after it is distributed in a noisy channel with excess noise $\Delta ^{2}(\hat{X}_{\epsilon})=\Delta ^{2}(\hat{Y}_{\epsilon})=\delta$ \cite{OneWayQin}. $\delta=0$ means that there is no excess noise, and only loss exists in the channel. $\delta>0$ means that there exists excess noise in the channel. So the covariance matrix of the CV entangled state \(\ket{\Psi}_{l}\) after distribution in the lossy or noisy channel is as following:

\begin{equation}
\sigma_{AB,\delta,\eta}=\left(\begin{array}{cccc}
V_{a} & 0 &  V_{c}  & 0 \\
0 & V_{a} & 0 & -V_{c} \\
V_{c}  & 0 & V_{b} & 0 \\
0 & -V_{c} & 0 & V_{b}
\end{array}
\right) =\left(
\begin{array}{cc}
V_{a} I & V_{c} Z \\
V_{c} Z & V_{b} I%
\end{array}%
\right) 
\end{equation}

with $V_{a}=\frac{V+V^{\prime}}{2}$, $V_{b}=\eta\frac{V+V^{\prime}}{2}+(1-\eta)(1+\delta)$ and $V_{c}=\sqrt{\eta} \frac{V^{\prime}-V}{2}$.  

\section*{APPENDIX B: Details of the Experiment} 

The Ti:sapphire laser (Coherent MBR-110) is about 1.6 GHz  blue detuned from $^{133}$Cs D1 line $6S_{1/2}, F=3\rightarrow6P_{1/2}, F'=4$ transition with a total power of 1.2 W. As shown in Fig. 5, the laser beam is split into two beams by a polarization beam splitter (PBS$_{1}$). The horizontally polarized beam is split into two beams by PBS$_{2}$ and they serve as the pump beams for two sets of four-wave mixing (FWM) processes in the same cesium vapor cell. The vertically polarized beam, with a power of 30 mW, passes through a resonance electro-optic modulator (EOM, Qubig GmbH PM - Cs) and three successive temperature-stabilized etalons to realize 9.2 GHz frequency red shift \cite{MaOL}. The OAM of topological charge $l=1$ or $l=2$ is added to the frequency red-shifted beam by passing it through a vortex phase plate (VPP, RPC Photonics). Then the frequency red-shifted beam is split into three beams, among which the first two beams serve as the probe beams of two FWM processes, and the third beam serves as the auxiliary beam for noisy channel. The pump and probe beams are combined in a Glan-laser (GL) polarizer and then cross each other in the center of the cesium vapor cell (Traid Technology Inc.) at an angle of 6 mrad. The vapor cell is 25 mm long and its temperature is stabilized at $103^{\circ}$C. 

The two sets of FWM processes are constructed in the same cesium vapor cell with a hight difference of 5 mm. The bottom FWM process is used to generate OAM multiplexed CV entangled state, while the top FWM process is used to generate spatially matched local oscillators (LOs) with the Pr and Conj fields. The pump power in the bottom FWM process for generating OAM multiplexed CV entanglement is 240 mW. The probe gain of the bottom FWM process is around 3, and the degree of initially generated CV entanglement is around $-$3.3$\pm$0.1 dB. The pump power and seed probe power of the top FWM processes are 450 mW and 100 $\mu$W, respectively, so that the shot-noise level (SNL) is around 10 dB higher than the electronic noise of the homodyne detector. The bottom FWM process is weakly seeded with a probe power of around 3 $\mu$W for relative phase locking of the Pr/Conj fields and their LOs in the balanced homodyne detections.

The lossy channel is simulated by a half-wave plate (HWP$_{10}$) and PBS$_{6}$. The noisy channel is modeled by combining the vertically polarized Pr field with an horizontally polarized auxiliary beam at PBS$_{5}$ followed by HWP$_{10}$ and PBS$_{6}$. The auxiliary beam carries the same frequency and topological charge with the Pr field, and is modulated by an amplitude modulator (AM) and a phase modulator (PM) with excess noise at 1.2 MHz . The amount of excess noise is adjusted by tuning the amplitude of the signal applied to the AM and PM, and evaluated in the units of SNL. For example, excess noise \(\delta=1\) corresponds to noise level that is 3 dB higher than the SNL. By tuning HWP$_{10}$, the lower transmission efficiency of the Pr field is, the higher excess noise is coupled to the Pr field. In practical quantum communication protocols, higher excess noise is coupled to quantum entangled state as the communication distance increases, accompanying lower transmission efficiency. Therefore, our experimental setting is similar to the realistic scenarios in practical noisy quantum channel. To characterize the OAM multiplexed CV entangled state, its covariance matrix is experimentally obtained by utilizing two sets of balanced homodyne detectors (BHDs, Thorlabs PDB450A). The interference visibilities for the two sets of BHDs are both around 99$\%$. The electrical gains of these two BHDs are both $10^5$ V/A. The original photodiodes are replaced by high quantum efficiency (QE) photodiodes with QE=98$\%$ at 895 nm (First sensor). To measure amplitude quadrature $\hat{X}$ or phase quadrature $\hat{P}$ of the Pr/Conj fields, the relative phases between them and their LOs are locked by applying feedback signal from proportional–integral–derivative circuits and high-voltage amplifiers to piezoelectric ceramics (PZT).

\section*{APPENDIX C: Measurement of the Covariance matrix }

To reconstruct covariance matrix of the CV quantum entangled state, we perform 6 different measurements on the output optical modes. These measurements include the variances of the amplitude and phase quadratures of the Conj field and Pr field $\Delta ^{2}\hat{X}_{-l,C}$, $\Delta ^{2}\hat{Y}_{-l,C}$, $\Delta ^{2}\hat{X}_{l,P}$, $\Delta ^{2}\hat{Y}_{l,P}$, as well as noise variances of their joint amplitude or phase quadrature $\Delta ^{2}(\hat{X}_{l,P}-\hat{X}_{-l,C}) $, and $\Delta ^{2}(\hat{Y}_{l,P}+\hat{Y}_{-l,C}) $, respectively. 
$\Delta ^{2}\hat{X}_{-l,C}$ and $\Delta ^{2}\hat{Y}_{-l,C}$ ($\Delta ^{2}\hat{X}_{l,P}$ and $\Delta ^{2}\hat{Y}_{l,P}$) are experimentally measured by locking the relative phase of Conj (Pr) field and its corresponding local oscillator of BHD$_{1}$ (BHD$_{2}$) at amplitude quadrature or phase quadrature. $\Delta ^{2}(\hat{X}_{l,P}-\hat{X}_{-l,C})$ and $\Delta ^{2}(\hat{Y}_{l,P}+\hat{Y}_{-l,C}) $ are experimentally measured by locking the relative phases of BHD$_{1}$ and BHD$_{2}$ at amplitude quadrature or phase quadrature, and then subtracting or adding the photocurrents with a radio-frequency subtractor or adder. The SNL is achieved by blocking the Pr and/or Conj field, so that only the noise of vacuum is measured. The settings of the spectrum analyzer (Agilent E4411B) are 30 kHz resolution bandwidth, 100 Hz video bandwidth, and zero span at 1.2 MHz. 

Fig. 6 shows the measured 6 noise variance levels of the CV entanglement of OAM multiplexed CV entangled state before being distributed in lossy and noisy channels ($\delta=0,\eta=1$). As shown in Fig. 6, the noise levels of amplitude and phase quadratures are $\Delta ^{2}\hat{X}_{-l,C}=\Delta ^{2}\hat{Y}_{-l,C}=\Delta ^{2}\hat{X}_{l,P}=\Delta ^{2}\hat{Y}_{l,P}$=3.6$\pm$0.1 dB, and the noise levels of correlated quadratures are $\Delta ^{2}(\hat{X}_{l,P}-\hat{X}_{-l,C}) $=$\Delta ^{2}(\hat{Y}_{l,P}+\hat{Y}_{-l,C}) $=$-$3.3$\pm$0.1 dB. Our FWM process works in the amplification regime, so amplitude quadratures and phase quadratures of the Pr/Conj fields show strong correlations and anti-correlations, respectively. It is clear that the degrees of entanglement for CV entangled states carrying topological charge \(l=1\), \(l=2\) are close to that of their spatially Gaussian counterpart \(l=0\).

With the measured 6 noise variances, the cross correlation matrix elements are calculated via 
\begin{align}
Cov( \hat{\xi}_{i},\hat{\xi}_{j}) & =\frac{1}{2}\left[ \Delta
^{2}\left( \hat{\xi}_{i}+\hat{\xi}_{j}\right) -\Delta ^{2}\hat{\xi}_{i}-\Delta ^{2}%
\hat{\xi}_{j}\right] , \\
Cov( \hat{\xi}_{i},\hat{\xi}_{j}) & =-\frac{1}{2}\left[ \Delta
^{2}\left( \hat{\xi}_{i}-\hat{\xi}_{j}\right) -\Delta ^{2}\hat{\xi}_{i}-\Delta ^{2}%
\hat{\xi}_{j}\right] .  \notag
\end{align}%

In the experiment, we obtain all the covariance matrices of quantum states with different transmission efficiencies and excess noise, and then calculate the smallest symplectic eigenvalue $\nu$ of the partially transposed covariance matrix, $\mathcal{G}^{A\rightarrow B}(\sigma _{AB})$ and $\mathcal{G}^{B\rightarrow A}(\sigma _{AB})$ to verify whether quantum entanglement and quantum steering exist.

\section*{APPENDIX D: Supplemental beam patterns}
Supplemental beam patterns of the OAM modes of the Pr and Conj beams generated from the FWM process are shown in Fig. 7. Pr and Conj beams carrying different topological charges $l=0$, $l=1$, and $l=2$ are generated by removing or switching different VPPs, and their beam patterns are taken after the GT polarizer with a camera. It is obvious that the dark hollow patterns of the Laguerre-Gaussian beams are enlarged as the topological charge increase. Furthermore, the beam patterns of the Pr and Conj fields are symmetric with respect to the pump beam.

\medskip
\noindent\textbf{Funding} National Natural Science Foundation of China (NSFC) (Grants No. 11974227, No. 11834010, No. 61905135, and No. 62005149); Fund for Shanxi ``1331 Project" Key Subjects Construction; Research Project Supported by Shanxi Scholarship Council of China (2021-003).

\medskip

\noindent\textbf{Disclosures.} The authors declare no conflicts of interest.
\medskip

$^{\dag }$These authors contributed equally to this Letter.


\begin{thebibliography}{99}
\bibitem{EPREntanglement} R. Horodecki, P. Horodecki, M. Horodecki, and K. Horodecki, ``Quantum entanglement," Rev. Mod. Phys. \textbf{81}, 865 (2009).

\bibitem{BraunsteinRMP} S. L. Braunstein and P. van Loock, ``Quantum information with continuous variables," Rev. Mod. Phys. \textbf{77}, 513 (2005).

\bibitem{KimbleQuanInt} H. J. Kimble, ``The quantum internet," Nature (London) \textbf{453}, 1023 (2008).

\bibitem{WeedbrookRMP} C. Weedbrook, S. Pirandola, R. Garc\'{i}a-Patr\'{o}n, N. J. Cerf, T. C. Ralph, J. H. Shapiro, and S. Lloyd, ``Gaussian quantum information," Rev. Mod. Phys. \textbf{84}, 621 (2012).

\bibitem{PhysicsReport} X. Wang, T. Hiroshima, A. Tomita, and M. Hayashi, ``Quantum information with Gaussian states," Physics Reports \textbf{448}, 1-111 (2007).

\bibitem{Bell} J. S. Bell, ``On the Einstein Podolsky Rosen paradox," Physics \textbf{1}, 195 (1964).

\bibitem{SteeringRMP} R. Uola, A. C. S. Costa, H. C. Nguyen, and O. G\"{u}hne, ``Quantum steering," Rev. Mod. Phys. \textbf{92}, 015001 (2020).

\bibitem{WisemanPRL} H. M. Wiseman, S. J. Jones, and A. C. Doherty, ``Steering, entanglement, nonlocality, and the Einstein-Podolsky-Rosen paradox," Phys. Rev. Lett. \textbf{98}, 140402 (2007).

\bibitem{OneWayNatPhot} V. H\"{a}ndchen, T. Eberle, S. Steinlechner, A. Samblowski, T. Franz, R. F. Werner, and R. Schnabel, ``Observation of one-way Einstein-Podolsky-Rosen steering," Nat. Photonics \textbf{6}, 596 (2012).

\bibitem{ANUexp} S. Armstrong, M. Wang, R. Y. Teh, Q. Gong, Q. He, J. Janousek, H. A. Bachor, M. D. Reid, and P. K. Lam, ``Multipartite Einstein-Podolsky-Rosen steering and genuine tripartite entanglement with optical networks," Nat. Phys. \textbf{11}, 167 (2015).

\bibitem{OneWayPryde} S. Wollmann, N. Walk, A. J. Bennet, H. M. Wiseman, and G. J. Pryde, ``Observation of genuine one-Way Einstein-Podolsky-Rosen steering," Phys. Rev. Lett. \textbf{116}, 160403 (2016).

\bibitem{OneWayGuo} K. Sun, X.-J. Ye, J.-S. Xu, X.-Y. Xu, J.-S. Tang, Y.-C. Wu, J.-L. Chen, C.-F. Li, and G.-C. Guo, ``Experimental quantification of asymmetric Einstein-Podolsky-Rosen steering," Phys. Rev. Lett. \textbf{116}, 160404 (2016).

\bibitem{OneWayQin} Z. Qin, X. Deng, C. Tian, M. Wang, X. Su, C. Xie, and K. Peng, ``Manipulating the direction of Einstein-Podolsky-Rosen steering," Phys. Rev. A \textbf{95}, 052114 (2017).

\bibitem{XiaoY2017} Y. Xiao, X.-J. Ye, K. Sun, J.-S. Xu, C.-F. Li, and G.-C. Guo, ``Demonstration of multisetting one-way Einstein-Podolsky-Rosen steering in two-qubit systems," Phys. Rev. Lett. \textbf{118}, 140404 (2017).

\bibitem{cvdv} A. Cavaill\`{e}s, H. Le Jeannic, J. Raskop, G. Guccione, D. Markham, E. Diamanti, M. D. Shaw, V. B. Verma, S. W. Nam, and J. Laurat, ``Demonstration of Einstein-Podolsky-Rosen steering using hybrid continuous- and discrete-variable entanglement of light," Phys. Rev. Lett. \textbf{121}, 170403 (2018).

\bibitem{WangPRL2020} M. Wang, Y. Xiang, H. Kang, D. Han, Y. Liu, Q. He, Q. Gong, X. Su, and K. Peng, ``Deterministic distribution of multipartite entanglement and steering in a quantum network by separable states," Phys. Rev. Lett. \textbf{125}, 260506 (2020).

\bibitem{SteeringQKD} C. Branciard, E. G. Cavalcanti, S. P. Walborn, V. Scarani, and H. M. Wiseman, ``One-sided device-independent quantum key distribution: Security, feasibility, and the connection with steering," Phys. Rev. A \textbf{85}, 010301(R) (2012).

\bibitem{SchnabelOneSidedQKD} T. Gehring, V. H\"{a}ndchen, J. Duhme, F. Furrer, T. Franz, C. Pacher, R. F. Werner, and R. Schnabel, ``Implementation of continuous-variable quantum key distribution with composable and one-sided-device-independent security against coherent attacks," Nat. Commun. \textbf{6}, 8795 (2015).

\bibitem{PKLamOneSidedQKD} N. Walk, S. Hosseini, J. Geng, O. Thearle, J. Y. Haw, S. Armstrong, S. M. Assad, J. Janousek, T. C. Ralph, T. Symul, H. M. Wiseman, and P. K. Lam, ``Experimental demonstration of Gaussian protocols for one-sided device-independent quantum key distribution," Optica \textbf{3}, 634 (2016).

\bibitem{WaveMultiplexing1} M. Pysher, Y. Miwa, R. Shahrokhshahi, R. Bloomer, and O. Pfister, ``Parallel generation of quadripartite cluster entanglement in the optical frequency comb," Phys. Rev. Lett. \textbf{107}, 030505 (2011).

\bibitem{WaveMultiplexing2} J. Roslund, R. M. de Ara\'{u}jo, S. Jiang, C. Fabre, and N. Treps, ``Wavelength-multiplexed quantum networks with ultrafast frequency combs," Nat. Photonics \textbf{8}, 109 (2014).

\bibitem{PolarMultiplexing} H. Liu, J. Wang, H. Ma, and S. Sun, ``Polarization-multiplexing-based measurement-device-independent quantum key distribution without phase reference calibration," Optica \textbf{5}, 902 (2018).

\bibitem{TimeMultiplexing1} S. Yokoyama, R. Ukai, S. C. Armstrong, C. Sornphiphatphong, T. Kaji, S. Suzuki, J. Yoshikawa, H. Yonezawa, N. C. Menicucci, and A. Furusawa, ``Ultra-large-scale continuous-variable cluster states multiplexed in the time domain," Nat. Photonics \textbf{7}, 982 (2013).

\bibitem{TimeMultiplexing2} A. Marandi, Z. Wang, K. Takata, R. L. Byer, and Y. Yamamoto, ``Network of time-multiplexed optical parametric oscillators as a coherent Ising machine," Nat. Photonics \textbf{8}, 937 (2014).

\bibitem{TimeMultiplexing3} M. Larsen, X. Guo, C. Breum, J. Neergaard-Nielsen, U. Andersen, ``Deterministic generation of a two-dimensional cluster state," Science \textbf{366}, 369 (2019).

\bibitem{SpatialMultiplexing1} D. J. Richardson, J. M. Fini, and L. E. Nelson, ``Space-division multiplexing in optical fibres," Nat. Photonics \textbf{7}, 354 (2013).

\bibitem{SpatialMultiplexing2} Y. Pu, Y. Wu, N. Jiang, W. Chang, C. Li, S. Zhang, and L. Duan, ``Experimental entanglement of 25 individually accessible atomic quantum interfaces," Sci. Adv. \textbf{4}, eaar3931 (2018).

\bibitem{AllenOAM}  L. Allen, M. W. Beijersbergen, R. J. C. Spreeuw, and J. P. Woerdman, ``Orbital angular momentum of light and the transformation of Laguerre-Gaussian laser modes," Phys. Rev. A \textbf{45}, 8185 (1992).

\bibitem{OAMMultiplexing1} J. Wang, J. Yang, I. M. Fazal, N. Ahmed, Y. Yan, H. Huang, Y. Ren, Y. Yue, S. Dolinar, M. Tur, and A. E. Willner, ``Terabit free-space data transmission employing orbital angular momentum multiplexing," Nat. Photonics \textbf{6}, 488 (2012).

\bibitem{OAMMultiplexing2} N. Bozinovic, Y. Yue, Y. Ren, M. Tur, P. Kristensen, H. Huang, A. E. Willner, and S. Ramachandran, ``Terabit-scale orbital angular momentum mode division multiplexing in fibers," Science \textbf{340}, 1545 (2013).

\bibitem{ZeilingerPNAS} R. Fickler, G. Campbell, B. Buchler, P. K. Lam, and A. Zeilinger, ``Quantum entanglement of angular momentum states with quantum numbers up to 10,010," Proc. Natl. Acad. Sci. U.S.A. \textbf{113}, 13642-13647 (2016).

\bibitem{Lu18modes} X. L. Wang, Y. H. Luo, H. L. Huang, M. C. Chen, Z. E. Su, C. Liu, C. Chen, W. Li, Y. Q. Fang, X. Jiang, J. Zhang, L. Li, N. L. Liu, C. Y. Lu, and J. W. Pan, ``18-qubit entanglement with six photons’ three degrees of freedom," Phys. Rev. Lett. \textbf{120}, 260502 (2018).

\bibitem{QinLight} Z. Qin, A. S. Prasad, T. Brannan, A. MacRae, A. Lezama, and A. I. Lvovsky, ``Complete temporal characterization of a single photon," Light Sci. Appl. \textbf{4}, e298 (2015).

\bibitem{QinPRL} Z. Qin, L. Cao, H. Wang, A. M. Marino, W. Zhang, and J. Jing, ``Experimental generation of multiple quantum correlated beams from hot rubidium vapor," Phys. Rev. Lett. \textbf{113}, 023602 (2014).

\bibitem{EntangledImages} V. Boyer, A. M. Marino, R. C. Pooser, and P. D. Lett, ``Entangled images from four-wave mixing," Science \textbf{321}, 544 (2008).

\bibitem{JingHexaEnt} K. Zhang, W. Wang, S. Liu, X. Pan, J. Du, Y. Lou, S. Yu, S. Lv, N. Treps, C. Fabre, and J. Jing, ``Reconfigurable hexapartite entanglement by spatially multiplexed four-wave mixing processes," Phys. Rev. Lett. \textbf{124}, 090501 (2020).

\bibitem{OAMFWM2008} A. M. Marino, V. Boyer, R. C. Pooser, P. D. Lett, K. Lemons, and K. M. Jones, ``Delocalized correlations in twin light beams with orbital angular momentum," Phys. Rev. Lett. \textbf{101}, 093602 (2008).

\bibitem{SU11} F. Hudelist, J. Kong, C. Liu, J. Jing, Z. Y. Ou, and W. Zhang, ``Quantum metrology with parametric amplifier-based photon correlation interferometers," Nat. Commun. \textbf{5}, 3049 (2014).

\bibitem{PooserOptica} R. C. Pooser and B. Lawrie, ``Ultrasensitive measurement of microcantilever displacement below the shot-noise limit," Optica \textbf{2}, 393 (2015).

\bibitem{JingBiOAM} X. Pan, S. Yu, Y. Zhou, K. Zhang, K. Zhang, S. Lv, S. Li, W. Wang, and J. Jing, ``Orbital-angular-momentum multiplexed continuous-variable entanglement from four-wave mixing in hot atomic vapor," Phys. Rev. Lett. \textbf{123}, 070506 (2019).

\bibitem{JingTriOAM} S. Li, X. Pan, Y. Ren, H. Liu, S. Yu, and J. Jing, ``Deterministic generation of orbital-angular-momentum multiplexed tripartite entanglement," Phys. Rev. Lett. \textbf{124}, 083605 (2020).

\bibitem{JingHexaOAM} W. Wang, K. Zhang, and J. Jing, ``Large-scale quantum network over 66 orbital angular momentum optical modes," Phys. Rev. Lett. \textbf{125}, 140501 (2020).

\bibitem{JingQuanTele} S. Liu, Y. Lou, and J. Jing, ``Orbital angular momentum multiplexed deterministic all-optical quantum teleportation," Nat. Commun. \textbf{11}, 3875 (2020).

\bibitem{OAMDistribution1} D. Cozzolino, E. Polino, M. Valeri, G. Carvacho, D. Bacco, N. Spagnolo, L. Oxenløwe, and F. Sciarrino, ``Air-core fiber distribution of hybrid vector vortex-polarization entangled states," Adv. photonics. \textbf{1}, 046005 (2019).

\bibitem{OAMDistribution2} G. Vallone, V. D’Ambrosio, A. Sponselli, S. Slussarenko, L. Marrucci, F. Sciarrino, and P. Villoresi, ``Free-space quantum key distribution by rotation-invariant twisted photons," Phys. Rev. Lett. \textbf{113}, 060503 (2014).

\bibitem{OAMDistribution3} Y. Chen, W. Shen, Z. Li, C. Hu, Z. Yan, Z. Jiao, J. Gao, M. Cao, K. Sun, and X. Jin, ``Underwater transmission of high-dimensional twisted photons over 55 meters," PhotoniX 1:5 (2020).

\bibitem{MaOL} R. Ma, W. Liu, Z. Qin, X. Su, X. Jia, J. Zhang, and J. Gao, ``Compact sub-kilohertz low-frequency quantum light source based on four-wave mixing in cesium vapor," Opt. Lett. 
\textbf{43}, 1243 (2018).

\bibitem{PPTCriterion} R. Simon, ``Peres-Horodecki separability criterion for continuous variable systems," Phys. Rev. Lett. \textbf{84}, 2726 (2000).

\bibitem{QuantificationPRL} I. Kogias, A. R. Lee, S. Ragy, and G. Adesso, ``Quantification of Gaussian quantum steering," Phys. Rev. Lett. \textbf{114}, 060403 (2015).

\bibitem{TopologicalCharge} P. Vaity, J. Banerji, and R. P. Singh, ``Measuring the topological charge of an optical vortex by using a tilted convex lens," Phys. Lett. A \textbf{377}, 1154 (2013).

\bibitem{npjDeng} X. Deng, Y. Liu, M. Wang, X. Su, and K. Peng, ``Sudden death and revival of Gaussian Einstein–Podolsky–Rosen steering in noisy channels," npj Quantum Inf. \textbf{7}, 65 (2021).

\bibitem{VectorBeam} Z. Zhu, M. Janasik, A. Fyffe, D. Hay, Y. Zhou, B. Kantor, T. Winder, R. W. Boyd, G. Leuchs, and Z. Shi, ``Compensation-free high-dimensional free-space optical communication using turbulence-resilient vector beam," Nat. Commun. \textbf{12}, 1666 (2021).

\end{thebibliography}
\end{document}